\documentclass[%
 aip,
 jmp,%
 amsmath,amssymb,
 reprint,%
]{revtex4-1}

\usepackage{graphicx}% Include figure files
\usepackage{dcolumn}% Align table columns on decimal point
\usepackage{bm}% bold math
\usepackage{natbib}
\begin{document}

\title[Aggregate Properties of Preimages]{Computing Aggregate Properties of Preimages for 2D Cellular Automata}

\author{Randall D. Beer}
 \email{rdbeer@indiana.edu.}
\affiliation{ 
Cognitive Science Program, Indiana University, Bloomington, IN 47406
}%

\date{\today}

\begin{abstract}
Computing properties of the set of precursors of a given configuration is a common problem underlying many important questions about cellular automata. Unfortunately, such computations quickly become intractable in dimension greater than one. This paper presents an algorithm --- incremental aggregation --- that can compute aggregate properties of the set of precursors exponentially faster than na{\"i}ve approaches. The incremental aggregation algorithm is demonstrated on two problems from the two-dimensional binary Game of Life cellular automaton: precursor count distributions and higher-order mean field theory coefficients. In both cases, incremental aggregation allows us to obtain new results that were previously beyond reach.
\end{abstract}

\maketitle

\begin{quotation}
%In addition, each article in Chaos is preceded by a lead paragraph targeted at non-specialist readers. This paragraph provides a sense of the context of the work and conveys the primary results in language that is accessible to the journal's broad interdisciplinary audience.

As simple models of distributed dynamical systems, cellular automata (CAs) have found wide application across many areas of science. A CA consists of a regular grid of cells, each of which can be in a finite number of states, that evolves in discrete steps according to some rule that assigns to each cell a new state based on its previous state and that of its neighbors. Perhaps the most famous such model is Conway's two-dimensional Game of Life cellular automaton. Many interesting questions about CAs can be framed as questions about the set of precursor configurations that can evolve into a given target configuration under the action of the update rule. Unfortunately, the set of precursors can be very expensive to compute. This paper presents a method for computing aggregate properties of the set of precursors of a configuration that is exponentially faster than previous approaches. The method is demonstrated on two problems from the Game of Life.

\end{quotation}

\section{Introduction}

Cellular automata are paradigmatic examples of spatiotemporal complex systems. They have a long history, dating from work by Ulam and von Neumann in the late 1940s and early 1950s. Cellular automata have been used to model a wide range of systems, including gases, fluids, excitable media, morphogenesis, tumor growth, the spread of forest fires, traffic flow, urban sprawl, and parallel computation.\cite{Chopard,KierSeybold,Worsch} In addition, they have served as an important testbed for exploring theoretical concepts such as pattern formation, criticality, emergence and computational universality.\cite{Hanson,DurandLose} 

Many interesting questions about a cellular automaton (CA) can be reduced to the computation of some property of the set of preimages of configurations of that automaton. The Garden of Eden problem asks for the configurations of a CA which have no precursors, that is, which can only occur as initial conditions. Finding the basin of an attractor involves tracing its precursors back to the Garden of Eden states that form its boundary. Statistical mechanical calculations require partition functions, which involve counting the number of precursors with a given property. And so on. Although practical methods exist for computing such properties for one-dimensional CAs,\cite{Jen,Wuensche} there has been much less progress on higher-dimensional CAs due to the $O(L^{n^d})$ scaling of na{\"i}ve approaches for an $L$-state size-$n$ CA in $d$ dimensions.

This paper describes a new algorithm --- incremental aggregation --- that can achieve an exponential speedup on problems involving aggregate properties of the entire set of preimages of a configuration in two-dimensional cellular automata. After some mathematical preliminaries, we present the algorithm and describe its efficient implementation. The algorithm is then demonstrated on two problems in Conway's Game of Life cellular automata.\cite{Conway,Adamatzky} First, we use it to count the number of precursors of individual configurations and to compute the distribution of precursor counts over the complete set of configurations of a given grid size. We show empirically that the algorithm achieves an $O(2^{n^2})$ to $O(2^n)$ reduction in execution time on precursor counting. Second, we compute the coefficients of the 3rd-order mean field theory for the Game of Life. Along the way, various optimizations that can be incorporated into the algorithm to further improve its performance are described. The paper concludes with a brief discussion of possible extensions to the algorithm and potential future applications.

\section{Preliminaries}

A {\em cellular automaton} is an autonomous dynamical system $\langle S^L_\Sigma, T, f^t \rangle$ consisting of an $L$-dimensional state space $S^L_\Sigma$ whose elements are drawn from an alphabet of symbols $\Sigma$, an ordered time set $T$ (typically ${\mathbb Z}^*$) and an evolution operator $f^t$ (with $t \in T$) defined as the $t$-times composition of a global transition function $f: S^L_\Sigma \rightarrow S^L_\Sigma$.  A key feature of cellular automata is that $f$ can be written element-wise in a uniform way, so that the global dynamics unfolds according to the same local law at all points in the space. Note that $f$ implicitly induces both a local and global topology on $S^L_\Sigma$ by virtue of the way in which the next state of one element of $S^L_\Sigma$ depends on other elements. The three most common global topologies are finite/bounded (the space has an edge beyond which all states are assumed to have some fixed value), finite/unbounded (the space is periodic in all directions) and infinite.

In the Game of Life (GoL) cellular automaton, $\Sigma=\{0,1\}$ and $S^L_\Sigma$ has the local topology of a rectangular grid of cells. The GoL update rule can be written element-wise as 
\begin{equation}
s^{t+1}_{x,y} = \delta_{M,3} + s^{t}_{x,y}\delta_{M,2},
\end{equation}
where $s^{t}_{x,y}$ is the state of the cell at location $(x,y)$ at time $t$, $M$ is the number of $1$ cells in that cell's Moore neighborhood (the eight cells surrounding it) and $\delta_{i,j}$ is the Kronecker delta function (which takes on the value $1$ when $i=j$ and the value $0$ otherwise). In words, a cell will be ON in the next time step if either exactly three of its neighbors are currently ON or it and exactly two of its neighbors are currently ON, otherwise it will be OFF. For notational simplicity only, we will assume in this paper that the rectangular grid is in fact square, with $L = n\times n$. We will also assume that the grid is sufficiently large that its global topology is irrelevant.

\begin{figure}[b]
\includegraphics[width=7cm]{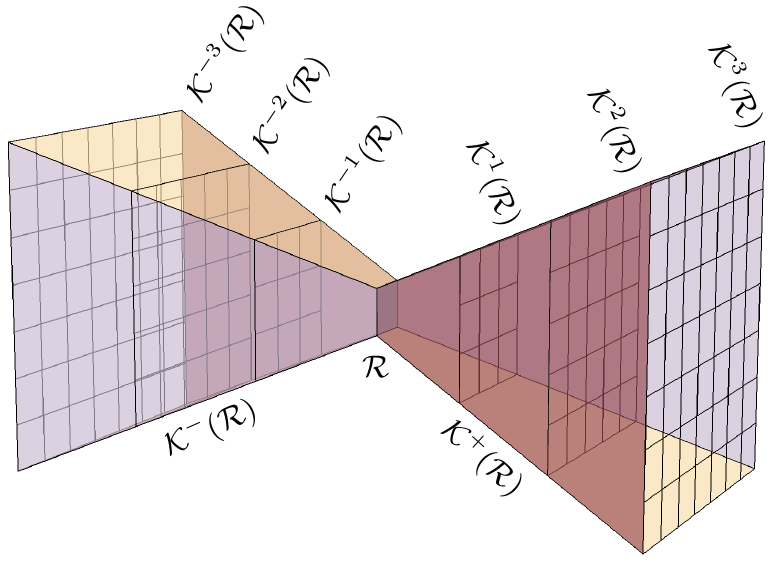}
\caption{The forward and backward light cones of a region $\mathcal{R}$ (here a single cell) in the Game of Life.}
\label{fig:lightcone}
\end{figure}

Consider some $n\times n$ region $\mathcal{R}$ of a 2-dimensional grid at time $t=0$. This region sits at the intersection of a forward light cone of cells that it can influence, $\mathcal{K}^+(\mathcal{R}) = \bigcup_t\mathcal{K}^t(\mathcal{R})$, extending into the future and a backward light cone of cells that can influence it, $\mathcal{K}^-(\mathcal{R}) = \bigcup_t\mathcal{K}^{-t}(\mathcal{R})$, extending into the past (Figure \ref{fig:lightcone}). A {\em configuration} of a region is an assignment of $0$ or $1$ to each of its constituent cells. If $C_{\mathcal{R}}$ is a particular configuration of $\mathcal{R}$, then its {\em preimage} $f^{-1}(C_{\mathcal{R}})$ is the set of all configurations $C'_{\mathcal{K}^{-1}(\mathcal{R})}$ of $\mathcal{K}^{-1}(\mathcal{R})$ such that $f_{\mathcal{R}}(C'_{\mathcal{K}^{-1}(\mathcal{R})})=C_{\mathcal{R}}$ (where $f_{\mathcal{R}}$ denotes $f$ with its codomain restricted to $\mathcal{R}$). This definition generalizes straightforwardly to the $t$-preimage
\begin{equation}
f^{-t}(C_\mathcal{R}) = \left\{C'_{\mathcal{K}^{-t}(\mathcal{R})} \left|  f^t_\mathcal{R}(C'_{\mathcal{K}^{-t}(\mathcal{R})}) = C_\mathcal{R}\right\}\right..
\end{equation}
Here, $f^t$ denotes the $t$-times composition of $f$ and $f^{-t}$ denotes the $t$-times inverse of $f$. Henceforth, we will drop the $\mathcal{R}$ subscripts from our notation, since the relevant regions are usually clear from context.

Our task in this paper is to efficiently compute a given aggregate property of the set of preimages of a configuration. Because this is an exponential problem,\cite{Sutner} the best that we can hope to do is make the exponent as small as possible. Accordingly, it will be useful to have some baseline preimage algorithms against which to judge our progress. The simplest such algorithm, which we will call the {\em forward algorithm}, is to iterate over all possible configurations $C'$ of $\mathcal{K}^{-t}(\mathcal{R})$, evolve each of them forward $t$ times, and then select those for which $f^t(C') = C$ for further processing. Since, due to the Moore neighborhood in GoL, effects spread at a speed of one cell per time, step the number of cells in $\mathcal{K}^{-t}(\mathcal{R})$ is $(n+2t)^2$. Thus, the running time of this algorithm scales as $O(2^{(n+2t)^2})$. Despite the fact that forward iteration of GoL can be made extremely fast, the scaling behavior of this forward algorithm severely limits its practical use.

A second approach, which we will call the {\em visitation algorithm}, is to work backward from a given target pattern, doing a depth-first search through the possible inversions $f^{-1}(s_{x,y})$ of each cell state $s_{x,y}$ in turn and identifying the globally consistent ones. In the best case, the visitation algorithm should scale as $O(|f^{-t}(C)|)$, although in the worst case it can degenerate to the $O(2^{(n+2t)^2})$ scaling of the forward algorithm. The visitation algorithm is about the best we can hope to do if we insist on ``touching'' each precursor individually.

\section{The Incremental Aggregation Algorithm}

\begin{figure}
{\centering\includegraphics[width=6cm]{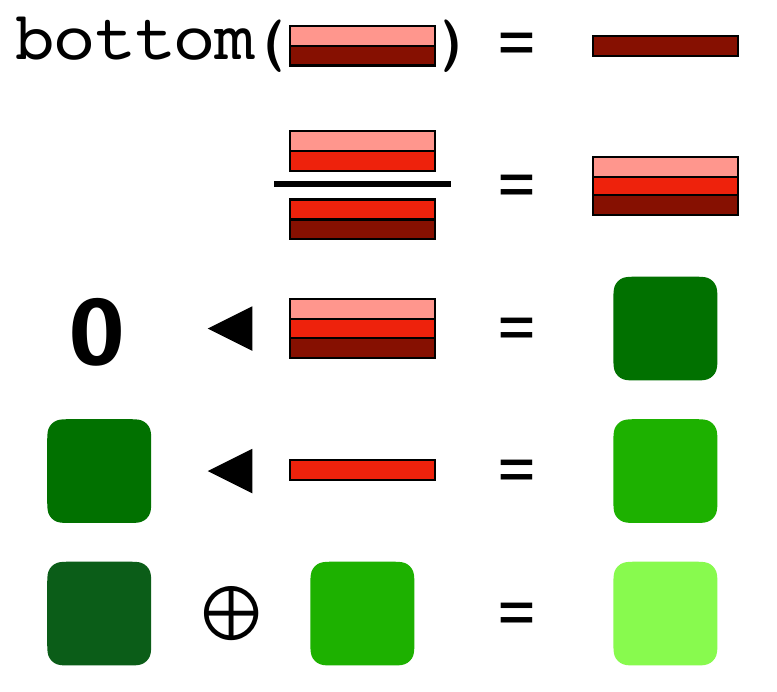}}
\caption{A graphical illustration of the key operations used in the incremental aggregation algorithm. Shaded bars represent precursor rows (with identical shading denoting identical contents), and rounded squares represent aggregate values (with brighter shades representing larger values).}
\label{fig:operators}
\end{figure}

\begin{figure*}[ht]
{\centering\includegraphics[width=15cm]{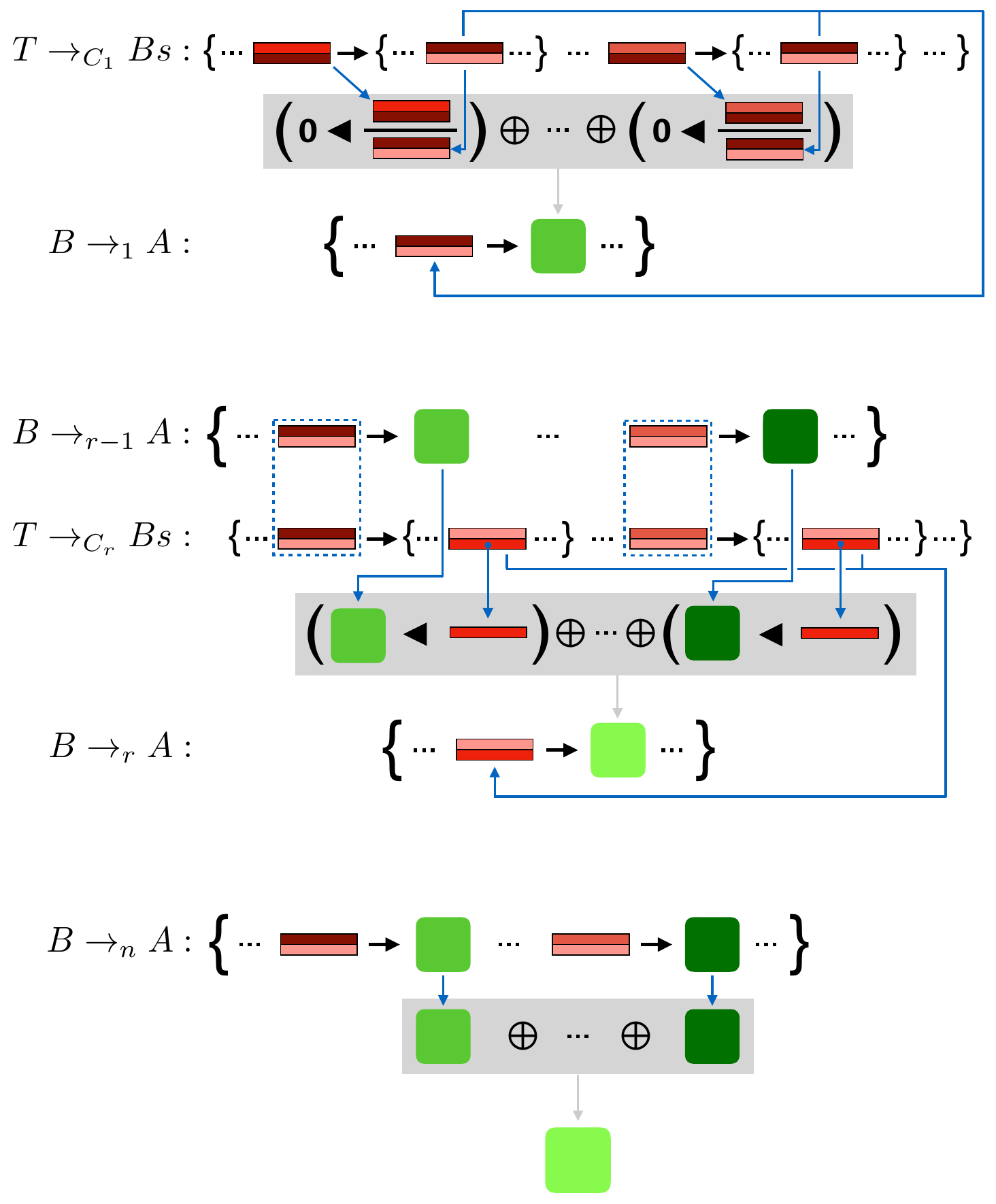}}
\caption{A graphical illustration of of the operation of the incremental aggregation algorithm. Conventions are the same as in Figure~\ref{fig:operators}. (Top) The initialization step. (Middle) The row extension step. (Bottom) The final aggregation step.}
\label{fig:IAA}
\end{figure*}

In many applications, we are only interested in aggregate properties of the entire set of preimages of a given configuration. For example, the number of such precursors is an aggregate property. In that case, we can often compute the desired property without explicitly visiting each precursor. The basic idea of the incremental aggregation algorithm is to work row-by-row, combining the desired aggregate property of each partial precursor with that of each matching set of new row precursors. Henceforth, our notation will assume a radius-1 CA such as GoL for simplicity, although the algorithm can easily be applied to larger radius CAs. Due to the level of abstraction at which the algorithm is formulated, while reading this section it may be useful to refer to the example of precursor counting described in the next section for a concrete example. Graphical illustrations of the key operators and steps of the algorithm are provided in Figures~\ref{fig:operators} and \ref{fig:IAA}, respectively.

Let $C$ be an $n \times n$ configuration and let $C'$ be one of its $(n+2)\times(n+2)$ precursors. Let $C_{r_1,r_2}$ denote the partial configuration consisting of rows $r_1$ through $r_2$ inclusive of $C$, with indices ranging from 1 to $n$. We will call $C_{1,s}$ the $s$-top of $C$ and $C_{n-s,n}$ its $s$-bottom. As a special case, $C_r$ denotes the $r$th row of $C$. We will sometimes need to treat this restriction as an operation so that, for example, ${\tt bottom}(C_{1,r})$ extracts the last row of $C_{1,r}$, i.e. $C_r$. A similar notation can be employed for $C'$ except that its row indices run from $0$ to $n+1$, so that a precursor of $C_1$ would be denoted $C'_{0,2}$. Finally, we denote by $\frac{t}{b}$ the overlapping vertical concatenation of a matching (i.e., ${\tt bottom}(t) = {\tt top}(b)$) pair of a 2-top $t$ and a 2-bottom $b$ to form the corresponding $3$-row configuration (Figure~\ref{fig:operators}).

Let $\langle\mathbf{0},\blacktriangleleft,\oplus\rangle$ be an aggregator for a particular domain of aggregation $\mathcal{A}$, consisting of an identity element $\mathbf{0}$, an extension operator $\blacktriangleleft$ which updates an $(r-1)$-partial aggregate value in light of a new row precursor, and an aggregation operator $\oplus$ which combines two aggregate values into one (Figure~\ref{fig:operators}). For example, for precursor counting in Section IV, the identity element will be the integer 0, the extension operator will increment a partial count each time a valid extension of a partial precursor is encountered, and the aggregation operator will add two partial counts together.

Let $B\rightarrow_r A = \{\ldots,b_i\rightarrow a_i,\ldots \}$ be a sequence of row-indexed maps from the 2-bottoms of the partial precursors of $C_{1,r}$ to $\mathcal{A}$. These maps index the partial aggregate values we have already computed up to row $r$ by their 2-bottoms for easy extension. We use the notation $M[\![x]\!]$ to refer to the value associated with the key $x$ in the map $M$ and the notation $M[\![x]\!]\leftarrow y$ to denote an assignment to that value. Note that if $b$ does not appear as a key in the bottom-to-aggregate map $B\rightarrow_r A$, the default value for $M[\![x]\!]$ is ${\mathbf 0}$, i.e., the identity element for the aggregator.

We assume the existence of a set of $C_r$-indexed maps $T \rightarrow_{C_r} Bs = \{\ldots, t_j \rightarrow\{\ldots, b_k, \ldots\}, \ldots \}$ from the 2-tops of all row precursors $C'_{r-1,r+1}$ of $C_r$ to the corresponding sets of matching 2-bottoms. These maps organize the precursors to row $C_r$ by their 2-tops, collecting all 2-bottoms that share the same 2-top. These top-to-bottom maps can be precomputed for any given grid width by either the forward precursor algorithm or the visitation algorithm. Note that a given 2-bottom $b_k$ can be associated with more than one 2-top $t_j$ because both 2-tops and 2-bottoms individually underdetermine the $C'_{r-1,r+1}$ from which they are derived.

The central task of the incremental aggregation algorithm is to compute $B\rightarrow_r A$ from $B\rightarrow_{r-1} A$ using $T\rightarrow_{C_r}Bs$. This row extension step proceeds as follows. For each pair of entries $(b_i\rightarrow a_i)\in(B\rightarrow_{r-1}A)$ and $(t_j\rightarrow\{\ldots,b'_k,\ldots\})\in(T\rightarrow_{C_r}Bs)$ for which $b_i = t_j$, we extend the $(r-1)$-partial aggregate value $a_i$ by computing $a_i \blacktriangleleft {\tt bottom}(b'_k)$. Then we aggregate this value with any previous value $a_k$ already associated with $b'_k$ in the new map $B\rightarrow_r A$ we are constructing, so that the entry for $b'_k$ in $B\rightarrow_r A$ becomes $a_k \oplus (a_i \blacktriangleleft {\tt bottom}(b'_k))$.

In order to initialize this process, we need the bottom-to-aggregate map $B\rightarrow_1A$ for the first row. This initial map could be computed easily if the set of precursors to $C_1$ were explicitly available, but it is not. However, we can reconstruct this set from the associated 2-tops and 2-bottoms in $T\rightarrow_{C_r}Bs$. Coupling this initialization step with a final aggregation over all partial results in $B\rightarrow_n A$ produces the incremental aggregation algorithm (Figure~\ref{fig:IAA}).\\

\noindent{\tt AggregateOverPrecursors}$(C,\langle\mathbf{0},\blacktriangleleft,\oplus\rangle)$\\[4pt]
\noindent\hspace*{5pt}----- {Initialize first row} -----\\
\noindent\hspace*{5pt}$(B\rightarrow_1 A)\leftarrow \emptyset$\\
\noindent\hspace*{5pt}{\bf for}~$(t\rightarrow bs)\in(T\rightarrow_{C_1} Bs)$\\
\noindent\hspace*{15pt}{\bf for}~$b\in bs$\\
\noindent\hspace*{25pt}$(B\rightarrow_1 A)[\![b]\!]\leftarrow (B\rightarrow_1 A)[\![b]\!]\oplus({\bf 0} \blacktriangleleft\frac{t}{b})$\\[4pt]
\noindent\hspace*{5pt}----- {Process each subsequent row} -----\\
\noindent\hspace*{5pt}{\bf for}~$r=2$~{\bf to}~$n$~{\bf do}\\
\noindent\hspace*{15pt}$(B\rightarrow_r A)\leftarrow \emptyset$\\
\noindent\hspace*{15pt}{\bf for}~$(b\rightarrow a)\in(B\rightarrow_{r-1} A)$\\
\noindent\hspace*{25pt}{\bf for}~$b'\in(T\rightarrow_{C_r}Bs')[\![b]\!]$\\
\noindent\hspace*{33pt}$(B\rightarrow_r A)[\![b']\!]\leftarrow(B\rightarrow_r A)[\![b']\!]\oplus(a\blacktriangleleft{\tt bottom}(b'))$\\[4pt]
\noindent\hspace*{5pt}----- {Return final aggregate} -----\\
\noindent\hspace*{5pt}{\em result}~$\leftarrow{\bf 0}$\\
\noindent\hspace*{5pt}{\bf for}~$(b\rightarrow a)\in (B\rightarrow_n A)$\\
\noindent\hspace*{15pt}{\em result}~$\leftarrow result\oplus a$\\
\noindent\hspace*{5pt}{\bf return}~{\em result}\\

In an efficient implementation of this algorithm, both the $B\rightarrow_r A$ maps and the $T\rightarrow_{C_r} Bs$ maps can be represented as hash tables. Although these hash tables can become large, three optimizations are possible. First, for any given configuration $C$ we need only construct the $T\rightarrow_{C_r} Bs$ maps for the distinct $C_r$ entries that actually appear in $C$. Second, once $B\rightarrow_{r-1} A$ has been used to construct $B\rightarrow_r A$, the former map is no longer needed and its storage can be reclaimed. Third, we can avoid constructing the final $B\rightarrow_n A$ map by aggregating directly into the result for the last row.

We next examine two applications of the incremental aggregation algorithm. All timings presented in this paper were performed using an optimized C++ implementation of the incremental aggregation algorithm running on a 3.0GHz 8-core/16-thread 2013 Apple Mac Pro (Xeon E5-1680 v2) with 64GB of RAM.

\section{Counting Precursors}

One of the simplest aggregate properties that we might want to compute is the  number of preimages of a configuration, $\left|f^{-1}(C)\right|$. For precursor counting, the aggregation domain $\mathcal{A}$ is ${\mathbb Z}^*$ and we use the aggregator $\langle 0,1+,+\rangle$, where the extension operator $1+$ increments its aggregate value argument and ignores its row precursor argument. Here the property we are aggregating is simply the existence of a precursor and we just sum this property to compute the total number of such precursors. If we have some number $a$ of $(r-1)$-partial precursors $C'_{0,r}$ with a given 2-bottom and we have an allowable extension of that partial precursor to an $r$-partial precursor $C'_{0,r+1}$, then we increment $a$ when we carry it over into the new map. In addition, since such extensions can come from many different $C'_{0,r}$, we must combine the new $a$ with any $a$ that already exists in the new map by summation. In effect, we are multiplying the number of each $C'_{0,r}$ with a given 2-bottom by the number of possible extensions of that $C'_{0,r}$ to $C'_{0,r+1}$. 

This observation leads to an optimization that can be applied to processing the final row of a target configuration. Not only can we avoid building $B\rightarrow_n A$, as mentioned above, but we can use multiplication to eliminate the innermost loop of processing for the last row. In particular, we terminate the row processing at row $n-1$ and then replace the final aggregation step with\\

\noindent\hspace*{5pt}----- {Return final count} -----\\
\noindent\hspace*{5pt}{\em result}~$\leftarrow 0$\\
\noindent\hspace*{5pt}{\bf for}~$(b\rightarrow a)\in (B\rightarrow_{n-1} A)$\\
\noindent\hspace*{15pt}{\em result}~$\leftarrow$~{\em result}~$+~a \times {\tt length}((T\rightarrow_{C_n} Bs')[\![b]\!])$\\
\noindent\hspace*{5pt}{\bf return}~{\em result}\\

\begin{figure}
{\centering\includegraphics[width=8.5cm]{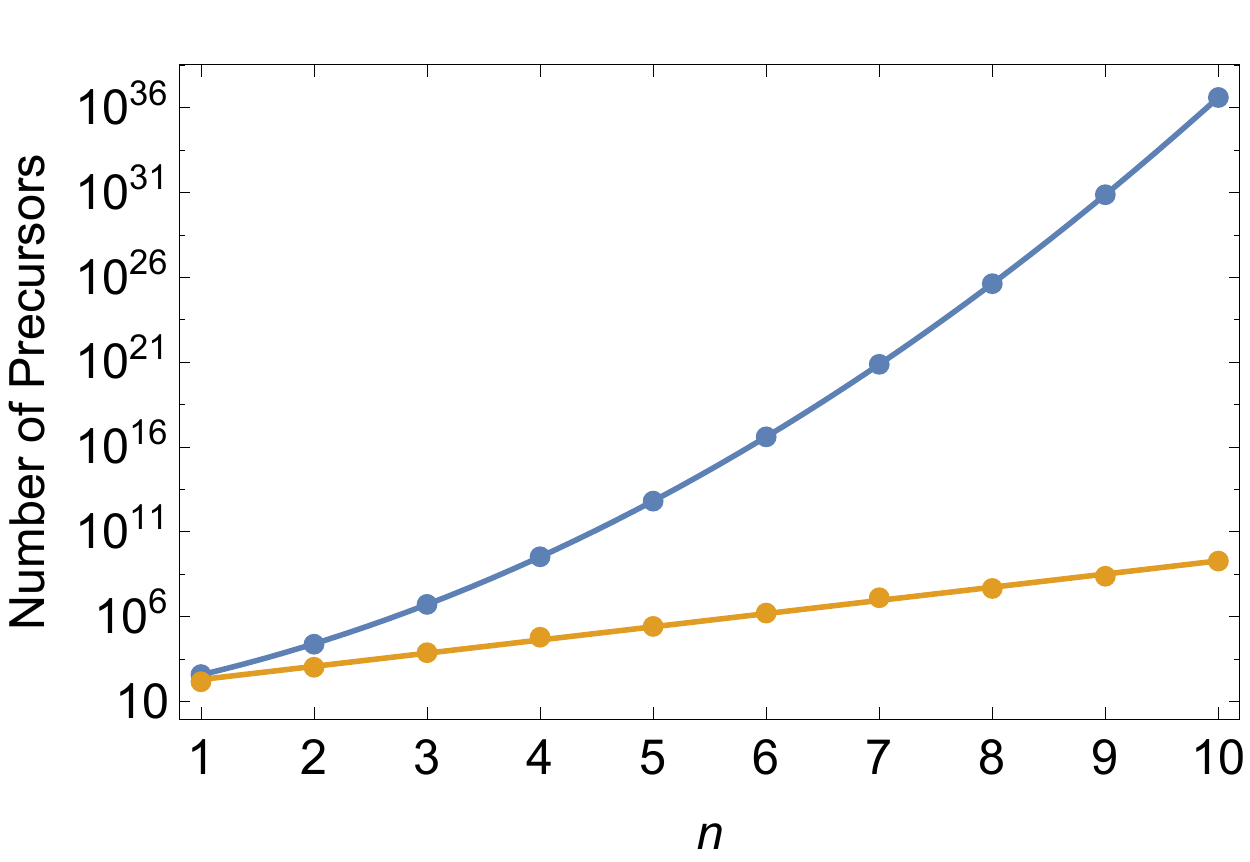}}
\caption{A log plot of the scaling of the number of precursors with grid size for $0^{n\times n}$ (blue) and $1^{n\times n}$ (yellow) in the Game of Life. The fits are $2^\wedge(0.810968 n^2 + 3.64563 n + 4.03293)$ and $2^\wedge(2.58318 n + 4.98204)$, respectively.}
\label{fig:countscaling}
\end{figure}

As an example of precursor counting, consider the family of $n \times n$ vacuum state (all $0$) configurations $0^{n\times n}$. Vacuum states are attractors of GoL and, since there is a strong tendency for a substantial fraction of a GoL grid to quickly decay to quiescence, the basins of attraction of vacuum states must be quite large. We can use the incremental aggregation algorithm to compute the 1-basin size of an $n \times n$ vacuum configuration by counting the number of $(n+2)\times(n+2)$ configurations that evolve to that configuration in one time step. As can be seen in Figure \ref{fig:countscaling}, this number grows very quickly with grid size (blue curve), with $\left|f^{-1}(0^{10\times 10})\right| = 3~965~375~048~845~134~539~385~175~457~630~019~267$. In contrast, the number of preimages of the family of all $1$ configurations $1^{n\times n}$ (which are not attractors but rather 1-precursors of the corresponding  vacuum state), grows much more slowly with grid size (yellow curve), reaching only $1~829~325~441$ at $n = 10$. 

\begin{figure}
{\centering\includegraphics[width=8.5cm]{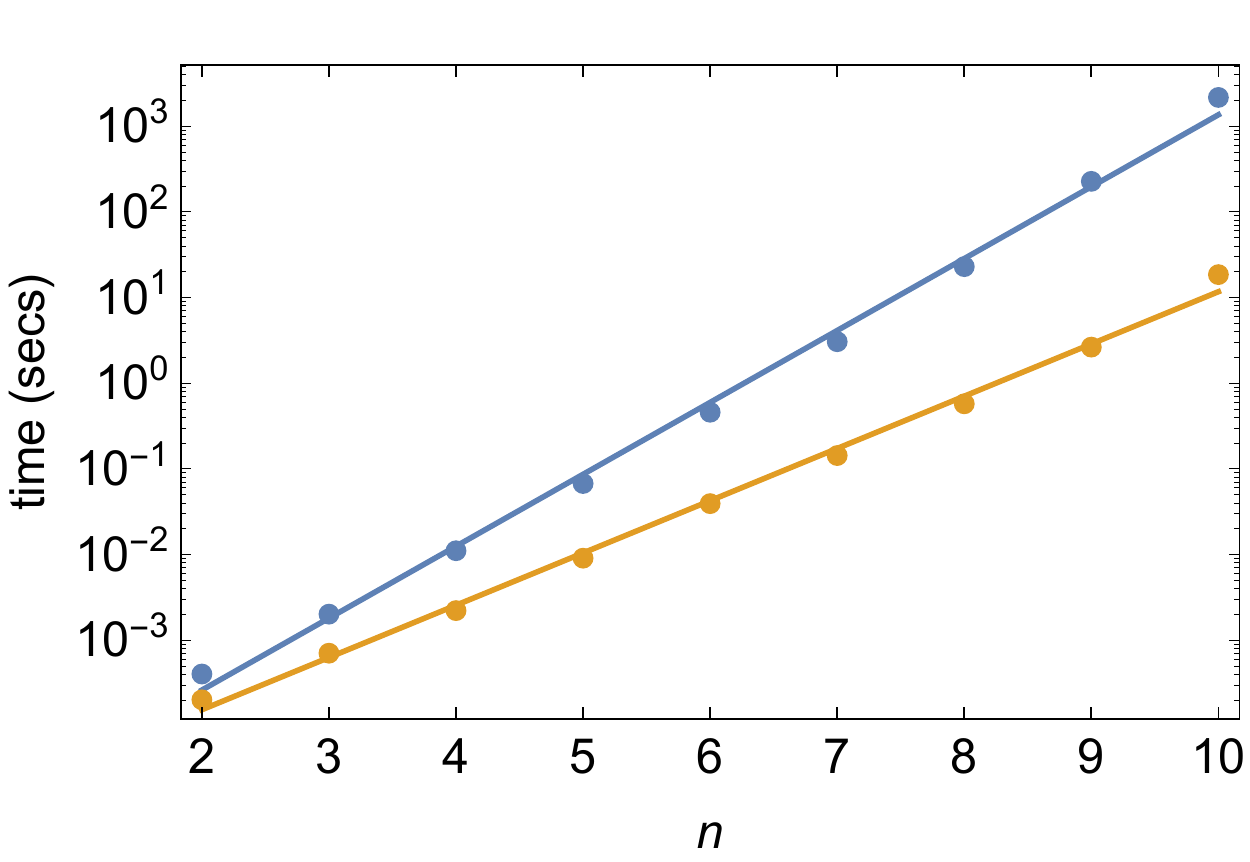}}
\caption{A log plot of the execution time scaling of precursor counting with grid size for $0^{n\times n}$ (blue) and $1^{n\times n}$ (yellow) in the Game of Life. The fits are $2^\wedge(2.78873 n - 17.4874)$ and $2^\wedge(2.02644 n - 16.726)$.}
\label{fig:timescaling}
\end{figure}

These two families of configurations also provide an excellent vehicle for probing the scaling behavior of the incremental aggregation algorithm. As shown in Figure \ref{fig:timescaling}, the time it takes this algorithm to compute such counts scales as $O(2^{c n})$, with $c\approx 2.79$ for $\left|f^{-1}(0^{n\times n})\right|$ and $c\approx 2.03$ for $\left|f^{-1}(1^{n\times n})\right|$. In both cases, the construction of the required $T\rightarrow_{C_r} Bs$ entries consumes a substantial fraction of this time. Fortunately, this cost can be amortized across multiple runs of the algorithm on configurations that share rows. Although still exponential, this scaling behavior is exponentially better than the $O(2^{(n+2)^2})$ scaling of the forward algorithm or the best case $O(\left|f^{-1}(0^{n\times n})\right|)\approx O(2^{0.8n^2})$ scaling of the visitation algorithm. Indeed, given the size of $\left|f^{-1}(0^{10\times 10})\right|$, an execution time of about 36 minutes is remarkable; it amounts to an effective speed of almost 2 decillion ($10^{33}$) configurations per second. Of course, the whole point of this algorithm is to avoid having to visit each precursor individually.

Now we ask a more ambitious question: What form does the full distribution of precursor counts take for a given grid size? Consider the $5\times5$ configurations. Even fully parallelized, a direct application of the incremental aggregation algorithm to the full set of $2^{25}$ $5\times5$ configurations takes well over nine hours. However, two additional optimizations are possible. First, we can take advantage of the $D_4$ symmetry of GoL to reduce the number of configurations whose precursors must be counted by roughly a factor of 8. If configurations are represented as $n^2$-bit integers, then these symmetry transformations can be efficiently implemented using bitwise techniques such as delta swaps.\cite{Knuth} Second, we can interleave the process of scanning through configurations with row-by-row aggregation. That is, for each possible configuration of the first row, we construct the corresponding $B\rightarrow_1 A$ map. For each of those, we then iterate over the possible configurations of the second row, constructing the corresponding $B\rightarrow_2 A$ maps, and so on until the last row. This ensures that each bottom-to-aggregate map will be constructed only once and used to its maximum effect before being discarded. With these
two optimizations, the execution time is reduced to just over 4 minutes

\begin{figure}
\includegraphics[width=8.5cm]{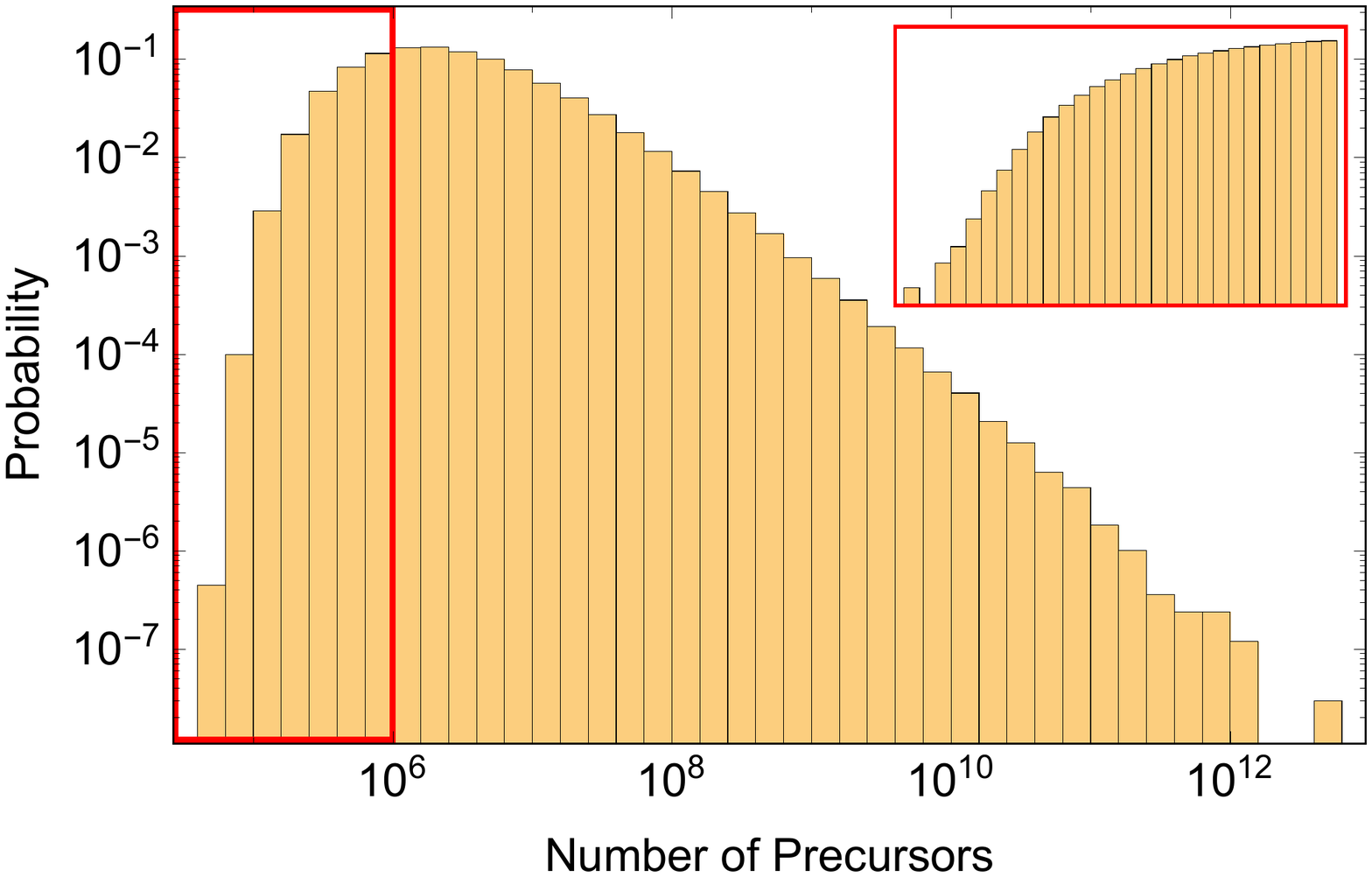}
\centering
\caption{A log-log plot of the distribution of precursor counts for $5\times 5$ grids in the Game of Life. The inset shows an expansion of the region indicated by the red rectangle.}
\label{fig:distribution}
\end{figure}

A log-log plot of the $5\times 5$ precursor count distribution is shown in Figure \ref{fig:distribution}. It is strongly skewed, with a power-law-like section on the right spanning over five orders of magnitude. Interestingly, while the largest precursor count ($6~272~107~655~484$) corresponds to $0^{5\times 5}$, the smallest precursor count ($40~528$) does not correspond to $1^{5\times5}$. Note the small gaps between the precursor counts for these two extreme configurations and the rest of the distribution, suggesting that they are somewhat exceptional. Smaller grids have a similar distribution, although it becomes considerably less smooth as grid size decreases. Monte Carlo samples of larger grids also suggest that their precursor count distributions take a similar form.

\section{Mean Field Theory}

\begin{figure}
\includegraphics[width=8cm]{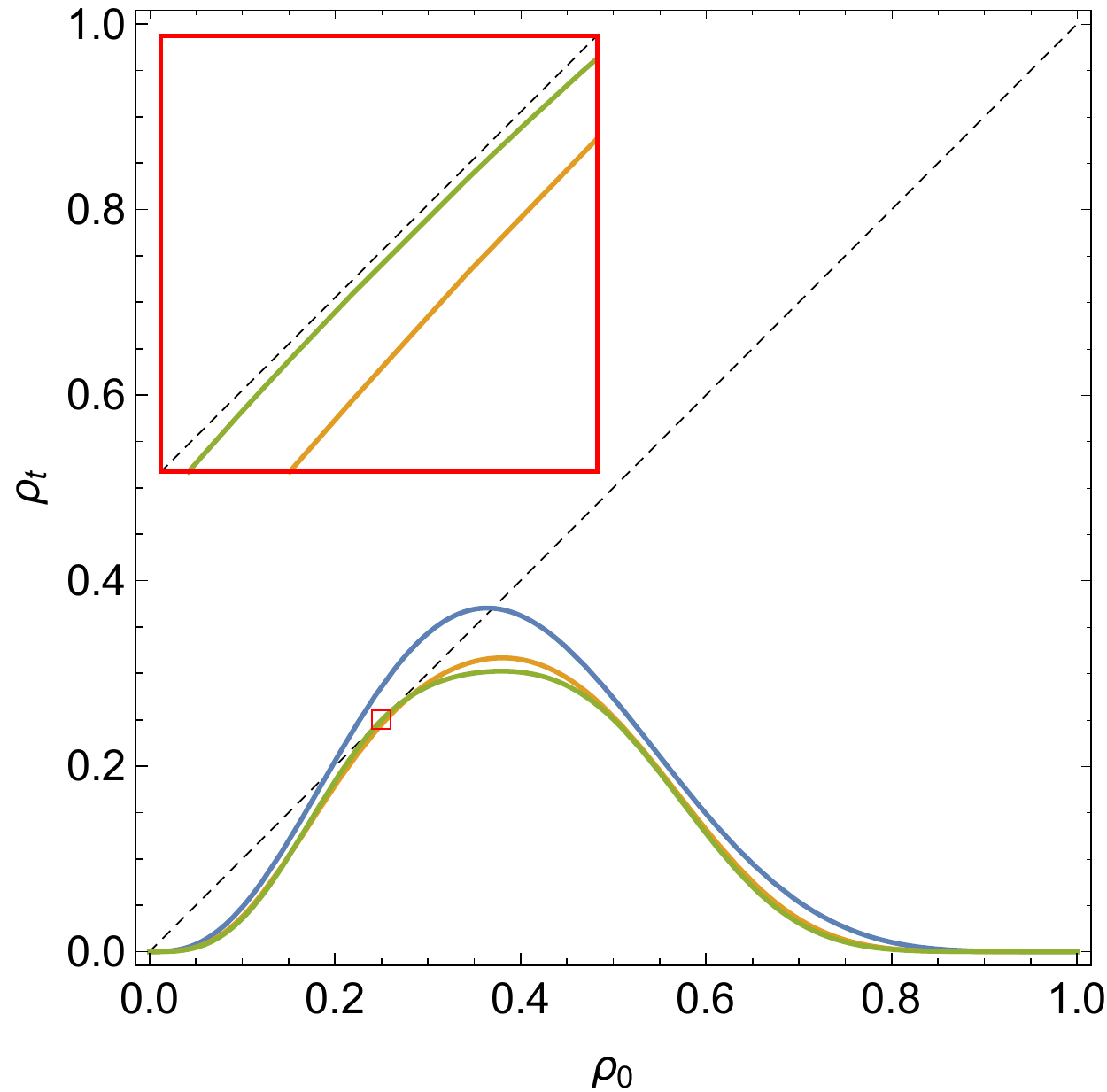}
\centering
\caption{Exact first- (blue), second- (yellow) and third- (green) order mean field theories $\rho_t(\rho_0)$ for the Game of Life. The inset shows an expansion of the region indicated by the small red box. Note that MFT3 is extremely close to tangency with the identity diagonal.}
\label{fig:MFT}
\end{figure}

More refined aggregate properties of precursors can also be computed by the incremental aggregation algorithm. Consider GoL mean field theory (MFT), which aims to predict the future grid density $\rho_t(\rho_0)$ from the initial density $\rho_0$. The first-order theory was derived by Schulman and Seiden \cite{SchulmanSeiden} (blue curve in Figure \ref{fig:MFT}). This theory predicts two stable fixed points at $\rho_\infty=0$ and $\rho_\infty\approx0.37017$ separated by an unstable fixed point at $\rho_\infty\approx0.19247$. Although the nontrivial fixed point predicted by MFT1 is observed under high noise conditions which serve to decorrelate the grid, experiments show that the asymptotic density of deterministic GoL is quite different: $\rho_\infty\approx0.029$. For this reason, Bagnoli et al. \cite{Bagnoli} derived an exact higher-order MFT that takes into account local correlations and they computed its second-order coefficients (yellow curve in Figure \ref{fig:MFT}). This theory predicts only a single stable fixed point at $\rho_\infty=0$ and it is quite useful for explaining several other aspects of the short-term evolution of GoL grids. Although even higher-order MFTs would provide a more accurate tool, Bagnoli et al. point out that computing their coefficients seems intractable. Next we show how the incremental aggregation algorithm can be used to efficiently compute the coefficients of the third-order theory MFT3. 

Higher-order MFTs can be derived as follows. Given the symmetries of GoL, the expected density after $t$ timesteps of a grid with initial density $\rho_0$ is equivalent to the expected state of an individual cell after $t$ time steps averaged over the set of $(2t+1)\times(2t+1)$ configurations with density $\rho_0$ that determine that state. Thus, if we define ${\left|f^{-1}(1)\right|}_k$ to be the Hamming weight decomposition of $\left|f^{-1}(1)\right|$ (that is, the number of distinct configurations in $f^{-1}(1)$ that contain exactly $k$ $1$s), then
\begin{equation}
\rho_t(\rho_0)={\mathbb E}\left[\frac{{\left|f^{-1}(1)\right|}_k}{\binom{(2t+1)^2}{k}}\right],
\end{equation}
\noindent where $k\sim B((2t+1)^2,\rho_0)$. Expanding this expectation using the PDF of the binomial distribution and cancelling the normalization factor, we obtain
\begin{equation}
\rho_t(\rho_0)=\sum_{k=0}^{(2t+1)^2} {\left|f^{-t}(1)\right|}_k(\rho_0)^k (1-\rho_0)^{(2t+1)^2 - k}.
\end{equation}

The $t$h-order MFT is thus a polynomial in $\rho_0$ defined by its coefficient vector $\mathbf{c}^t$ with components $c_k^t={\left|f^{-t}(1)\right|}_k$. This vector can be computed in three steps. First, the visitation algorithm is applied recursively to determine each symmetrically-distinct configuration $C\in f^{-(t-1)}(1)$ and its multiplicity. Then we use the incremental aggregation algorithm to compute the coefficient vector of each such $C$. Finally, we scale these vectors by their corresponding multiplicities and sum them to obtain $\mathbf{c}^t$.

For the computation of MFT coefficients, the aggregation domain is the space of coefficient vectors ${\mathbb Z}^{(2t+1)^2+1}$ and the aggregator is $\langle {\mathbf 0},\Rrightarrow,+\rangle$. Here, ${\mathbf 0}$ is the $0$ vector of length $(2t+1)^2+1$ and $+$ is vector addition. The extension operator $\Rrightarrow$ updates the coefficient vector for the $(r-1)$-partial precursor in light of a new row precursor by shifting the vector right by the Hamming weight of the new precursor row. The shift is necessary due to the positional nature of the coefficient vector representation. If coefficient $c_i$ in $\mathbf{c}$ has a value $v$ for an $(r-1)$-partial precursor and that partial precursors has a valid extension with a Hamming weight of $w$, then coefficient $c_{i+w}$ will have value $v$ in the resulting $r$-partial precursor. Note that, for this to work, ${\mathbf 0} \Rrightarrow x$ must be defined to set the component of ${\mathbf 0}$ corresponding to the Hamming weight of $x$ to $1$.

For MFT3, there are $1~065~921$ symmetrically-distinct $5\times5$ configurations in $f^{-2}(1)$ and $\mathbf{c}^3$ can be calculated in just over two hours (Table \ref{table:coefficients}). As Bagnoli et al. argued, the first three and at least the last five coefficients are necessarily $0$ due to the nature of the GoL update rule. The resulting $\rho_3(\rho_0)$ curve is shown in green in Figure \ref{fig:MFT}. It is similar to the second-order curve (yellow), but differs from it in two small but important ways. First, its peak is lower. In fact, simulations suggest that, as the order increases, the maximum of $\rho_t(\rho_0)$ monotonically decreases toward its asymptotic value of $\rho_\infty\approx0.029$. Second, as the inset to Figure \ref{fig:MFT} shows, MFT3 is much closer to tangency with the identity (diagonal line) than MFT2. This is relevant to debates about the criticality of GoL, which seem to have concluded that it is slightly subcritical. \cite{Bak,BlokBergersen,Reia} Simulations suggest that the distance to tangency steadily increases for higher orders, so MFT3 appears to be exceptional in this regard.

\begin{table}
\begin{tabular}{r r p{1.4cm} r r}
\hline
\hline
\noalign{\vskip 0.5ex}
$k$ & ${\left|f^{-3}(1)\right|}_k$ & & $k$ & ${\left|f^{-3}(1)\right|}_k$\\[0.5ex]
\hline
\noalign{\vskip 1ex}
 0 &                    0 & & 25 & ~~17~214~918~812~506\\
 1 &                    0 & & 26 & ~~13~959~267~158~216\\
 2 &                    0 & & 27 & ~~10~005~027~074~788\\
 3 &                   22 & & 28 &    6~324~472~595~876\\
 4 &                1~536 & & 29 &    3~517~402~910~448\\
 5 &               39~108 & & 30 &    1~716~624~960~572\\
 6 &              589~784 & & 31 &      732~951~761~550\\
 7 &            6~263~018 & & 32 &      272~636~042~324\\
 8 &           50~775~840 & & 33 &       87~784~126~868\\
 9 &          326~351~948 & & 34 &       24~243~942~928\\
10 &        1~697~457~928 & & 35 &        5~673~241~468\\
11 &        7~266~374~286 & & 36 &        1~106~668~596\\
12 &       25~954~288~768 & & 37 &          175~901~352\\
13 &       78~717~133~326 & & 38 &           22~028~188\\
14 &      207~045~941~204 & & 39 &            2~063~116\\
15 &      482~662~094~024 & & 40 &              132~588\\
16 &    1~016~038~322~780 & & 41 &                4~970\\
17 &    1~958~277~712~284 & & 42 &                   72\\
18 &    3~486~429~885~960 & & 43 &                    0\\
19 &    5~756~899~411~052 & & 44 &                    0\\
20 &    8~803~153~176~008 & & 45 &                    0\\
21 & ~~12~378~403~927~070 & & 46 &                    0\\
22 & ~~15~841~205~401~436 & & 47 &                    0\\
23 & ~~18~260~658~140~634 & & 48 &                    0\\
24 & ~~18~807~170~916~072 & & 49 &                    0\\ [1ex]
\hline
\hline
\end{tabular}
\caption{Coefficients ${\left|f^{-3}(1)\right|}_k$ of the third-order MFT $\rho_3(\rho_0)$.}
\label{table:coefficients}
\end{table}

What are the prospects for computing even higher-order MFT coefficients? Even the number of symmetrically-distinct $(2t+1)\times(2t+1)$ configurations in $f^{-t}(1)$ grows extremely quickly with $t$. Aside from this scaling, the most time-consuming portion of the computation is the dynamic memory allocation and deallocation of coefficient vectors in the $B\rightarrow_rA$ hash tables. This could probably be reduced significantly by using a more specialized statically-allocated data structure, by using a per-thread memory allocator, or by splitting the computation across multiple processes rather than multiple threads in order to avoid memory allocation contention.

\section{Conclusions}

This paper has presented a general method, the incremental aggregation algorithm, for computing aggregate properties of the set of preimages of a configuration in 2-dimensional cellular automata. This algorithm is exponentially faster than standard approaches because it scales with the number of rows rather than grid area. The algorithm was demonstrated on two problems in the Game of Life: precursor counting and coefficient calculation for higher-order mean field theories.

The incremental aggregation algorithm can be extended in many ways. With the appropriate aggregator it could be applied to other properties, such as finding Garden of Eden states or computing the basins of attraction of a given configuration to a desired depth. In addition, since nothing in the algorithm is specific to the Game of Life, it could also be applied easily to other cellular automata rules. For example, it would be interesting to see how the mean field theory of other ``Life-like'' CAs \cite{Adamatzky} compares to that for GoL. Finally, with some tedious but straightforward work, the algorithm could be applied to non-square and even non-rectangular grids and to CAs with more states, more dimensions, larger radii and other topologies. As always, the main limitation is the scaling of the problem relative to the available computational resources.

\begin{acknowledgments}
This paper has benefited greatly from discussions with Alexander Gates.
\end{acknowledgments}

\medskip

\bibliography{bibliography}

\end{document}